\pdfoutput=1 
\documentclass{JINST}

\title{Instrumental systematics and weak gravitational lensing}

\author{Rachel Mandelbaum$^a$\\
\llap{$^a$}McWilliams Center for Cosmology, Department of
  Physics, Carnegie Mellon University\\
5000 Forbes Ave., Pittsburgh, PA 15213, USA\\
E-mail: \email{rmandelb@andrew.cmu.edu}}

\abstract{We present a pedagogical review of the weak gravitational lensing measurement process and
  its connection to major scientific questions such as dark matter and dark energy.  Then we
  describe common ways of parametrizing systematic errors and understanding how they affect weak
  lensing measurements.  Finally, we discuss several instrumental systematics and how they fit into
  this context, and conclude with some future perspective on how progress can be made in
  understanding the impact of instrumental systematics on weak lensing measurements.}

\keywords{Data analysis; Image processing; Photon detectors for UV, visible and IR photons}

\begin{document}

\section{Motivation}

In our current cosmological model, only 4\% of the Universe consists of the baryonic matter from which stars and planets are
made.  To explain a wide
variety of cosmological observations, we have been forced to posit the existence of dark matter
(detected through its gravitational attraction) and dark energy (which causes a
repulsion that is driving the accelerated expansion of the Universe).  While we infer the existence
of these dark components, the question of what they actually are remains a mystery. 

Gravitational lensing
\cite{schneider06,2008ARNPS..58...99H,2010RPPh...73h6901M}
is the deflection of light from distant objects by the matter along its path to us -- all of the
matter, including dark matter.  Lensing measurements are thus directly sensitive to dark matter.
They also allow us to infer the properties of dark energy \cite{2002PhRvD..65b3003H}, because the
accelerated expansion of the Universe that it causes directly opposes the effects of gravity, which
tends to cause matter to clump into ever larger structures.  This measurement relies on the
small but spatially coherent distortions (known as weak shears) in the shapes of distant galaxies,
which provide a statistical map of cosmological large-scale structure along the line-of-sight.  A
key problem for weak lensing measurement is that there are many other effects in astronomical images
that cause coherent distortions of galaxy shapes and that must be effectively removed in order to
reliably measure weak lensing.

Because of its sensitivity to dark matter and dark energy, ever-larger surveys have been planned to
measure weak lensing increasingly precisely, including
Euclid\footnote{\texttt{http://sci.esa.int/euclid/}, \texttt{http://www.euclid-ec.org}}
\cite{2011arXiv1110.3193L}, LSST\footnote{\texttt{http://www.lsst.org/lsst/}}
\cite{2009arXiv0912.0201L}, and WFIRST-AFTA\footnote{\texttt{http://wfirst.gsfc.nasa.gov}}
\cite{2013arXiv1305.5422S}, all of which are Stage IV dark energy experiments according to the Dark
Energy Task Force \cite{2006astro.ph..9591A} definitions.  The very high signal-to-noise ratio
of the weak lensing measurements in these surveys will necessitate excellent understanding and control of systematic
uncertainties.

Broadly speaking, among the sources of systematic uncertainty in weak lensing measurements are (1)
instrumental systematics, (2) the need to remove the point-spread function (PSF) which blurs the
galaxy images to robustly infer the lensing shear, (3) the need to estimate redshifts for the
galaxies or at least statistically understand their distribution, and (4) theoretical uncertainties
such as the impact of baryons on the measured quantities (usually predicted using dark-matter-only
simulations) and failures in the assumption that galaxy shapes are randomly oriented in the absence of lensing.
Of these, issues (2) through (4)
 have been thoroughly explored in the literature (see, for example,
\cite{2006MNRAS.368.1323H,2006ApJ...636...21M,2007MNRAS.376...13M,2010MNRAS.401.1399B,2010MNRAS.405.2044B,2012MNRAS.423.3163K,2014arXiv1412.1825M,2014MNRAS.445.3382M,2014arXiv1407.6990T}
and references therein).  Instrumental systematics 
have received somewhat less attention, and are the main topic of discussion in this
pedagogical review.

We begin in section~\ref{sec:background} by discussing how weak lensing measurements are made,
and how those measurements are affected by systematic errors in general.  In section~\ref{sec:instr}
we discuss instrumental systematics more specifically.  We conclude in section~\ref{sec:concl} with future outlook.

\section{Scientific background}\label{sec:background}

\subsection{Weak lensing measurements}

In this subsection, we ignore the question of systematics, and give a basic summary of how weak
gravitational lensing measurements are carried out.

Cosmic shear, the measurement of weak lensing by large-scale structure, relies on correlating pairs of
galaxy shapes as a function of their separation on the sky.  Galaxies that are nearby on the sky 
tend to have correlated shapes due to having been lensed by associated structures
that subject them to similar weak lensing shears.  Galaxies that are farther apart on the sky 
have much less correlated weak lensing shears.

To carry out this measurement, the most basic requirement is a catalog of galaxy positions and their
shape estimates\footnote{Technically, one need not have a per-galaxy shape or shear estimate to
  reconstruct the lensing shear field; see \cite{2014MNRAS.438.1880B} for options that avoid calculating a
  per-object shear in favor of ensemble estimates.  However, these methods are
  relatively new and their use for cosmic shear requires further practical and theoretical
  development, so for the remainder of this review we assume that weak lensing measurements will use
  per-object galaxy shapes.}, which we take as a proxy for shear (but see section~\ref{subsec:sys}
for more details on the difference between galaxy shapes and shear).  While galaxy light profiles do
not in general have elliptical isophotes, they are nonetheless commonly described in terms of some
effective ellipticity magnitude $|\varepsilon$|, which can be defined in terms of the
minor-to-major axis ratio, and position angle $\phi$ with respect to some coordinate
system on the sky.  Alternatively, one can write two components of the ellipticity as $\varepsilon_1
= |\varepsilon|\cos{2\phi}, \varepsilon_2=|\varepsilon|\sin{2\phi}$, and define a complex
ellipticity as $\varepsilon=\varepsilon_1+\mathrm{i}\varepsilon_2$.  These trigonometric functions
of $2\phi$ are very common in mathematical expressions related to galaxy shapes or lensing shears,
due to their being 
spin-2 quantities (ellipses are invariant under $180^\circ$ rotations).  For more details on common
definitions of galaxy shapes, see for example \cite{2002AJ....123..583B}.

The galaxy shapes will have
some large random intrinsic components which is a key source of statistical error in most weak
lensing measurements, along with a small but fortunately coherent weak lensing shear.  The intrinsic
shape then averages out of correlation functions of galaxy shapes, while the lensing shear does not
(but see \cite{2014arXiv1407.6990T} for issues related to this basic assumption).  An estimate of the galaxy
redshift distribution is also needed in order 
to make a theoretical prediction for the cosmic shear correlations.  However, even more promising is
the so-called cosmic shear tomography, which involves dividing the galaxies into redshift bins based on
a per-galaxy photometric redshift estimate (see, for example, figure~1 in \cite{2013MNRAS.432.2433H}).  With $>1$ redshift bin, additional information can be
obtained by correlating galaxies both {\em within} and {\em across} those redshift slices, and
seeing how the shear signal changes in each case.

A typical cosmic shear correlation function estimator, for the correlation function corresponding to
redshift slices $i$ and $j$ is \cite{2013MNRAS.432.2433H}
\begin{equation}
\label{eq:xi}
\hat{\xi}_{\pm}^{ij}(\theta) = \frac{\sum w_a w_b \left[
    \varepsilon_t^i(\mathbf{x}_a)\varepsilon_t^i(\mathbf{x}_b) \pm 
\varepsilon_\times^i(\mathbf{x}_a)\varepsilon_\times^i(\mathbf{x}_b) \right]}{\sum w_a w_b }.
\end{equation}
The index $a$ runs over all galaxies in redshift slice $i$, while the index $b$ runs over all
galaxies in redshift slice $j$.  Galaxy pairs $(a,b)$ are identified, and their angular separation
on the sky $\theta$ (calculated using their positions $\mathbf{x}_{a,b}$) is used to put them into
bins in $\theta$.  Each galaxy has a weight factor $w_{a,b}$ that relates to uncertainty in the
shear estimates.  Each galaxy's shear estimate $\varepsilon$, which has two
components, is rotated into a coordinate system defined by the vector connecting galaxies $a$ and
$b$.  The component corresponding to orientation along or at $90^\circ$ with respect to that vector
is $\varepsilon_t$, and the component corresponding to orientation along the $\pm 45^\circ$ rotated
system is $\varepsilon_\times$.  The two estimated correlation functions for each pair of
redshift bins $i$ and $j$, $\hat{\xi}_{\pm}^{ij}(\theta)$, both relate to integrals over the matter
power spectrum along the line-of-sight.  For examples of recent cosmic shear measurements using several
different datasets,
 see \cite{2013MNRAS.432.2433H,2012ApJ...761...15L,2013ApJ...765...74J,2014MNRAS.440.1322H}.  These have been used to constrain
the amplitude of the matter power spectrum and its evolution with time, which tells us about the
equation of state of dark energy.

Another type of weak lensing measurement is cluster-galaxy or galaxy-galaxy lensing, which involves
identifying specific foreground galaxies or clusters that act as ``lenses'', and correlating them
with the positions of background sources.  This measurement is therefore a (lens) position
vs.\ (source) shape cross-correlation function, which makes it less subject to certain 
systematic errors in shear estimates.  It provides an estimate of the projected total matter surface
density around the chosen lenses.  While it has primarily been
used to make the connection between galaxies or clusters and their host dark matter halos (e.g.,
very recently, \cite{2015arXiv150202867C,2015arXiv150201024S}),
it can also be used in combination with estimates of galaxy clustering to constrain the growth of
cosmic structure and therefore dark energy \cite{2013MNRAS.432.1544M}, and to test whether General Relativity is the
effective theory of gravity on cosmological scales \cite{2010Natur.464..256R,2007PhRvL..99n1302Z}.

\begin{figure}[tbp]
\centering
\includegraphics[width=.44\textwidth]{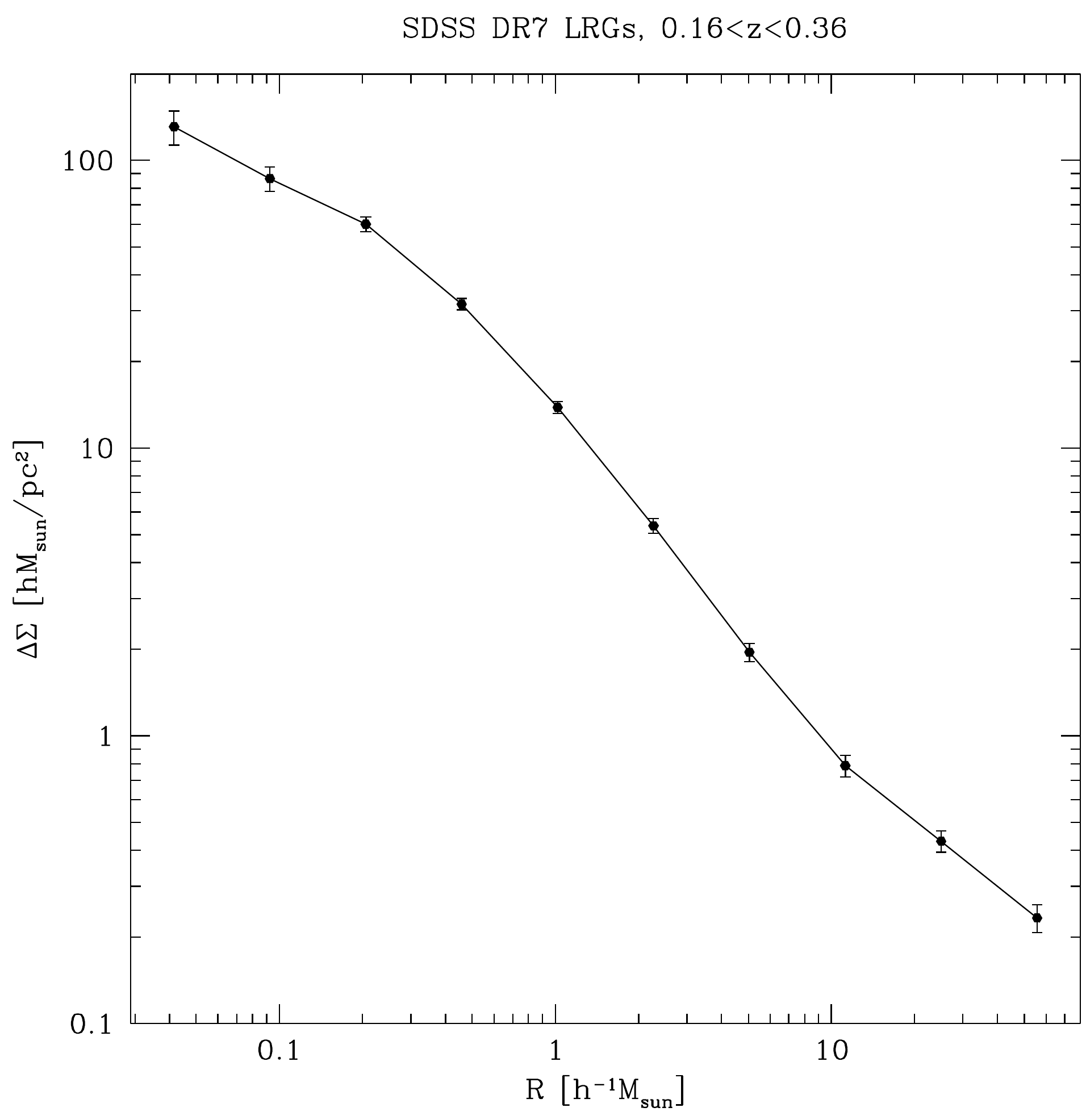}
\includegraphics[width=.54\textwidth]{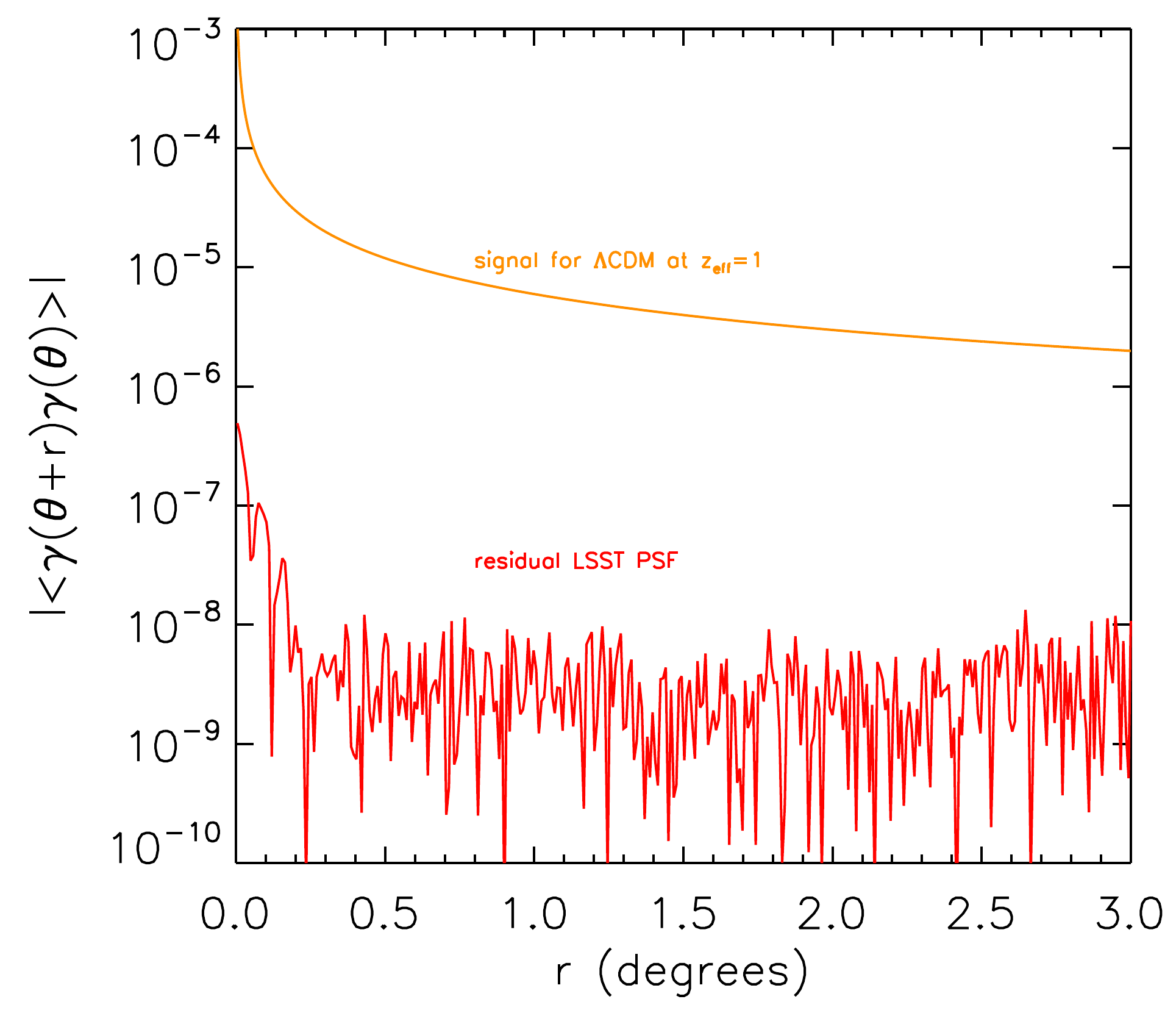}
\caption{{\em Left:} A galaxy-galaxy lensing measurement from the Sloan Digital Sky Survey (SDSS, \cite{2013MNRAS.432.1544M})
illustrating the projected mass profile around Luminous Red Galaxies (LRGs), including the host dark
matter halo below $1h^{-1}$Mpc and contributions from large-scale structure.  At the largest scales,
the signal corresponds to shears (shape distortions) of $\sim 5\times 10^{-4}$. {\em Right:} A theoretical cosmic shear
correlation function (figure 14.4 from \cite{2009arXiv0912.0201L}), illustrating the typical magnitude and scaling of the shear correlation
function  with separation, as well as an expected signal due to one
particular systematic error.
}
\label{fig:signals}
\end{figure}

Examples of both types of lensing observables are shown in figure~\ref{fig:signals}.

\subsection{The impact of systematics in shear estimation}\label{subsec:sys}

We now consider the main sources of systematics in estimating the lensing shears, how they are
commonly described, and how they affect the measured statistics used for many cosmic shear studies
in equation~\ref{eq:xi}.

First, it is important to bear in mind that the concept of a galaxy shape is not very well-defined.
Even moderate-resolution imaging reveals that galaxies do not have elliptical isophotes, but rather
may have some degree of clumpiness in their light profiles and/or ellipticity gradients (e.g. a
rounder bulge component in the inner region and a more extended and flattened disk).  Thus,
different methods of estimating galaxy shapes will often give quite different answers for the same 
galaxies.  Even the same method may give a different answer for a single galaxy when measuring in
different passbands, due to different components of the galaxy having different spectral energy
distributions.  These differences are, from a weak lensing perspective, not of interest.   
What is important is only that a given shape estimator have
a well-defined response {\em on average} to lensing shears, such that we can use the ensemble
averages to
measure statistics of the lensing shear field.

\begin{figure}[tbp]
\centering
\includegraphics[width=\textwidth]{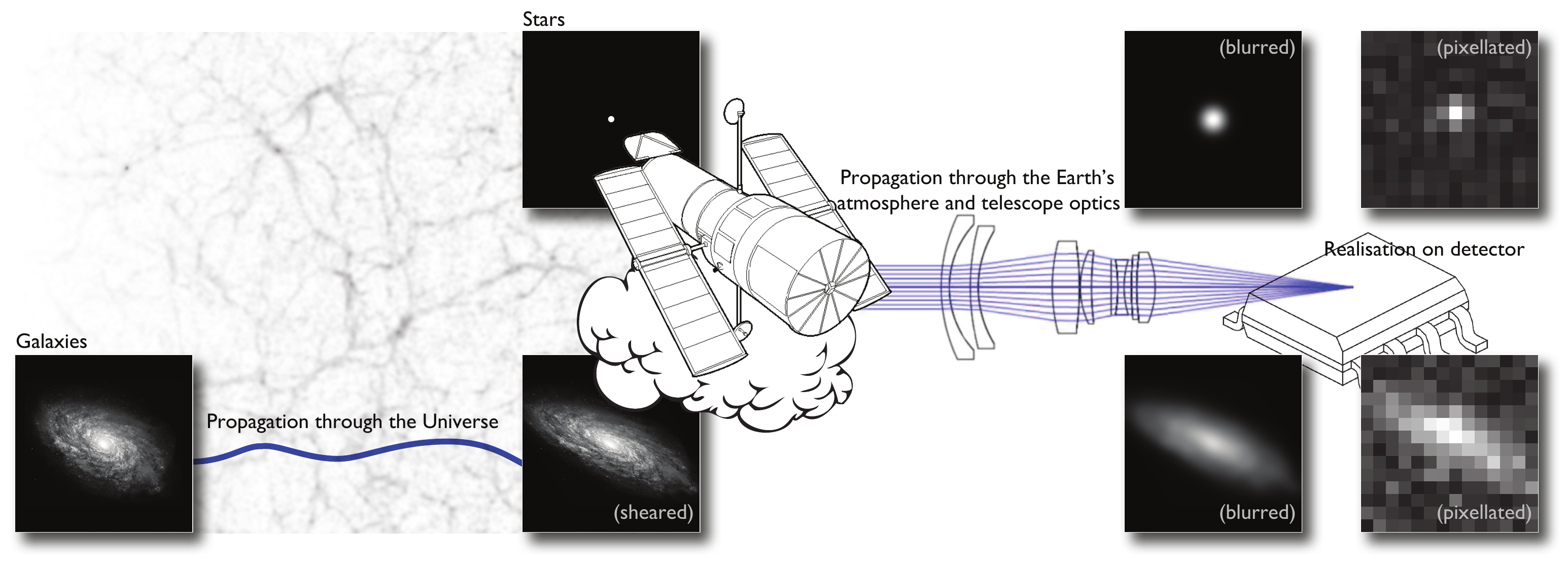}
\caption{An illustration of the process of
  gravitational lensing and other effects that change the apparent
  shapes of galaxies in the astronomical imaging process (from \cite{2014ApJS..212....5M}).
Instrumental systematics are not shown, but would appear after the PSF convolution.}
\label{fig:forward_G3}
\end{figure}

Images from telescopes do not directly show us the shapes of galaxies.  The galaxy light profiles
are modified by many effects.  First and foremost is the point-spread function (PSF) of the
atmosphere (for ground-based telescopes) and the telescope optics.  This can be modeled as a
convolution kernel that rounds galaxy shapes by a significant factor, but can have some
small ellipticity that gets coherently imprinted into galaxy shapes and mimics a cosmic shear
signal.  There are also distortions of coordinate systems that can be treated as affine
transformations rather than convolutions.  Images contain defects such as cosmic rays, bleed trails,
and others.  Finally, the images have noise, which results in significant biases in estimates of
shear \cite{2012MNRAS.427.2711K,2012MNRAS.425.1951R}.  A schematic illustrating the image generation process is shown
in figure~\ref{fig:forward_G3}.  Note that the process of removing the PSF to infer the
shears robustly has been the subject of focused attention from the weak lensing community for nearly
two decades (with several large community-wide blind challenges
\cite{2006MNRAS.368.1323H,2007MNRAS.376...13M,2010MNRAS.405.2044B,2012MNRAS.423.3163K,2014arXiv1412.1825M}).
Nonetheless, some issues remain to be solved, even when ignoring instrumental systematics.

A common description for systematic errors in shear estimates is a linear model \cite{2006MNRAS.366..101H},
\begin{equation}
\label{eq:bias}
\hat{\gamma} = (1+m)\gamma + c,
\end{equation}
which can be written separately for each of the two components of the shear. 
Here $\gamma$ is the true lensing shear, $\hat{\gamma}$ is the estimated shear, $m$ is the
multiplicative bias and $c$ the additive bias in the estimated shears. This additive bias is
in general linearly proportional to the PSF ellipticity \cite{2014arXiv1412.1825M}, though selection
effects that correlate with the PSF direction might cause a deviation from this typical relationship. Ideally, $m=c=0$.

\begin{figure}[tbp]
\centering
\includegraphics[width=\textwidth]{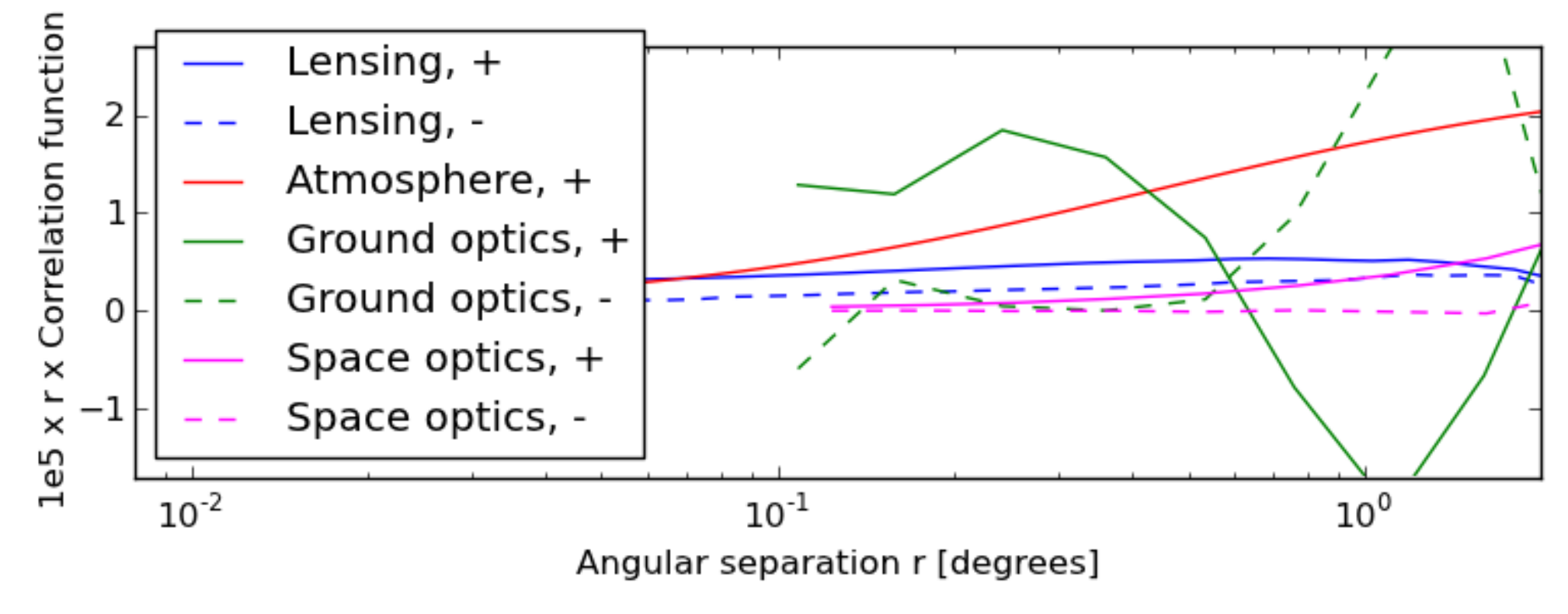}
\caption{Lensing shear correlation functions $\xi_\pm$, along with the analogous shape correlations
  for PSF anisotropies from several sources: typical atmospheric PSFs, and space- and ground-like
  optical PSFs.  As shown these can have quite different scalings with separation, and at some
  scales they have
  amplitudes above the cosmic shear signal, highlighting the need for very accurate removal.  From \cite{2014ApJS..212....5M}.}
\label{fig:cf}
\end{figure}

We can model galaxy-galaxy or cluster-galaxy lensing as a galaxy position $g$ vs.\ estimated shear
$\hat{\gamma}$ cross-correlation, or (schematically)
\begin{equation}
\langle g\hat{\gamma}\rangle = (1+m)\langle g\gamma\rangle + \langle g c\rangle.
\end{equation}
The first term on the right-hand side is the ideal signal with a multiplicative calibration bias determined by $m$.  The second, additive term relates
to the correlation between galaxy positions and additive systematics, which is typically either zero
or removable through simple means \cite{2005MNRAS.361.1287M}.  Using similar notation, cosmic shear can be
written as
\begin{equation}
\langle \hat{\gamma}\hat{\gamma}\rangle = (1+m)^2\langle \gamma\gamma\rangle + 2(1+m)\langle
c\gamma\rangle + \langle c c\rangle.
\end{equation}
The first term on the right-hand side is the real cosmic shear with a calibration factor $(1+m)^2\approx 1+2m$ (since 
typically $m\ll 1$).  The second term would involve a correlation
between additive systematics and the true shear, which may be nonzero by chance for small regions of
a survey but naturally averages to zero when considering large areas.  Finally, the last term is a
coherent additive term in the shear field, which can typically be diagnosed using correlations between the
galaxy shear estimates and star shapes \cite{2012MNRAS.427..146H}. Generally the scaling of this term
with separation will depend on the correlation function of the PSF ellipticity or other source of
additive systematics; for examples,
see figure~\ref{fig:cf}.  Since there are clear diagnostics for additive
systematics, the weak lensing community has primarily focused on  multiplicative biases in the 
recent past due to the difficulty in getting an absolute calibration.

In general, this linear model for systematic errors can be used to describe basic errors in shear
estimation due to the process of correcting galaxy shapes for the effects of the PSF.  For other
systematics, including instrumental ones, its
applicability may be limited.  For example, if the PSF is incorrectly modeled in a way that results
in the PSF being systematically too large by the same amount at all places on the sky, then this
could be modeled as a spatially constant $\langle m\rangle$ over the whole galaxy population.  Real errors in the modeling of the PSF tend to
have some scale-dependence, and one has to consider the full spatial correlation function of those
PSF modeling errors in order to really understand the impact on science.  The linear model is still
useful for providing basic intuition, however.

It is important to note that in general,
$m=m(S/N, \mathrm{size}, \mathrm{morphology})$ and therefore it is also effectively a function of
redshift.  The same is true for additive errors ($c$).  Thus, in a tomographic measurement, each
redshift slice must have its own effective systematic errors estimated and accounted for, rather
than treating the entire galaxy population in a monolithic way.

\section{Instrumental systematics}\label{sec:instr}

Instrumental systematics include a variety of effects such as nonlinearity, charge-transfer
inefficiency (most notably in the {\em Hubble Space Telescope} Advanced Camera for Surveys,
\cite{2010MNRAS.401..371M}), tree rings and edge distortions \cite{2014PASP..126..750P}, the
brighter-fatter effect \cite{2014JInst...9C3048A}, image defects leading to coherent selection
effects (masking bias, \cite{2014MNRAS.440.1296H}), amplifier overshoot, crosstalk, fringes, and
gate diffraction.  At a lower level, there are also possible chromatic instrumental effects, which
would affect galaxy and star images differently due to their different spectral energy
distributions.  Several of these effects (tree rings, edge distortions) can be treated as part of
the WCS (world coordinate system) mapping from image to world coordinates, albeit a rather complex
part of the WCS.  They do not necessarily
fit cleanly into the model for systematics presented in equation~\ref{eq:bias}.  Here we discuss two
of these issues in more detail.

\subsection{Brighter-fatter effect}

The brighter-fatter effect \cite{2014JInst...9C3048A,2015A&A...575A..41G} is a result of charge building up in pixels
and 
repelling additional electrons (pushing them toward neighboring pixels).  This makes the PSF larger for
brighter objects and smaller for fainter objects.  The problem for weak lensing is that bright stars
(signal-to-noise ratios of $\gtrsim 50$) are typically used to estimate the
PSF, while the galaxies used to estimate weak lensing shear are on average quite faint.  Thus, their
effective PSF will be smaller than the effective PSF model inferred from stars, and the PSF
correction routine will systematically over-correct for the blurring effects of the PSF, resulting
in overestimates of shear.

We can use a simple model for how the PSF affects galaxy shear measurements to estimate the impact
of the brighter-fatter effect (if left uncorrected).  Let us assume that for some particular CCD, the
linear PSF size (e.g., its FWHM) will be overestimated by 1\% due to the brighter-fatter effect (with 2\% being the maximum effect
seen in \cite{2014JInst...9C3048A} using a wide range of fluxes).
\cite{2004MNRAS.353..529H} showed that in the Gaussian approximation, the systematic bias in the
inferred shear ($m=\delta \hat{\gamma}/\hat{\gamma}$) due to a misestimate of the PSF size is
\begin{equation}
\frac{\delta \hat{\gamma}}{\hat{\gamma}} = \left(R_2^{-1}-1\right) \frac{\delta T^{(P)}}{T^{(P)}}.
\end{equation}
Here, $T^{(P)}$ is the trace of the second moment matrix of the PSF, and $\delta T^{(P)}/T^{(P)}$ is
its fractional error.  If the linear size of the PSF is misestimated by 1\% then $\delta
T^{(P)}/T^{(P)}\approx 0.02$.  $R_2$ is the galaxy ``resolution factor'', which is $0$ for
unresolved objects and approaches $1$ for highly resolved galaxies.  A galaxy near the resolution
limit might have $R_2\approx 0.25$, which results in a shear calibration bias $m=0.06$.  Moderately
resolved galaxies ($R_2\approx 0.5$) would have $m=0.02$.  For context, large future surveys require
that the calibration bias be reduced to the level of $m\lesssim(1-3)\times 10^{-3}$, or at least be
known and reliably corrected at that level \cite{2009arXiv0912.0201L,2013MNRAS.429..661M}, in order for the
calibration bias to not contaminate the dark energy constraints at a level comparable to the
statistical errors.

If the brighter-fatter effect has some directionality, then the PSF ellipticity for typical galaxies
might likewise be misestimated.  For typical methods of shear estimation, a few percent of the PSF ellipticity
leaks coherently as an additive $c$ term in the galaxy shears  \cite{2014arXiv1412.1825M}.  If the PSF ellipticity error due to
the brighter-fatter effect is $0.01$ (coherently), then a typical shear estimation method will end
up with a $c$ term of $\sim 3\times 10^{-4}$, again slightly above the requirements for future surveys.

 Clearly both the multiplicative and additive biases that may be induced by the brighter-fatter
 effect are larger than the requirements for upcoming surveys, so this effect must be well-modeled and
removed.

\subsection{Nonlinearity}

An example of a long-known instrumental effect is CCD nonlinearity, which can be important at the
high flux levels observed in the bright star images commonly used to estimate the PSF (but not for
most galaxy images).
CCD nonlinearity is commonly measured in the lab and then removed, making it an essentially ignored
effect for most weak lensing shear studies.  However, if there is some error in the nonlinearity
estimation, or if it is a poorly-tracked function of time such that the original corrections cannot
be used later on, this
can cause errors in the ``corrected'' images of bright stars used to estimate the PSF, meaning
that the PSF model used to correct the galaxy shapes is wrong.

An example of this error in practice has been seen in the SDSS.  As shown in \cite{2014MNRAS.440.1296H}, there is an
error in the PSF model and shear estimates in one of the six $r$-band CCDs, for which the most likely
explanation is an error in the nonlinearity corrections\footnote{R. Lupton, private
  communication.}. The impact  of this effect can be seen in
figure 8 of that work, which shows a
coherent systematic error in the galaxy shapes at the level of $5\times 10^{-3}$.  \cite{2013MNRAS.432.1544M} 
showed that the majority of the additive systematic errors in the lensing shear in the entire SDSS
area are eliminated after removing the $1/6$ of the data from that one CCD. 
While it seems unlikely that the nonlinearity for CCDs for future surveys will be misestimated in
advance, if there is any change in that nonlinearity in the years needed for these surveys to be
completed, then systematic errors of this type are possible.

\section{Conclusions}\label{sec:concl}

Instrumental systematics can leak into weak gravitational lensing measurements in
surprising ways.  We have described the quantities that are measured
as part of a weak lensing measurement, and gave examples for how two particular instrumental
systematics can affect those measurements.  In future, more work will be needed to fully
characterize the various instrumental systematics that will affect surveys such as the LSST.  There
are, fortunately, several means for doing so.  Laboratory tests can be used to understand the
magnitude 
of the effects, which can then be included in survey image simulation software such as that for LSST (PhoSim: \cite{2010SPIE.7738E..1OC,2013ascl.soft07011P}).  Several instrumental effects have also been included in
GalSim\footnote{\texttt{https://github.com/GalSim-developers/GalSim}} \cite{2014arXiv1407.7676R}, an
open-source image simulation software package, which can be used for more focused studies
of particular instrumental effects without including all possible systematics.  In either case, these simulation
packages can be used to simulate images and therefore estimate the effects of these systematics on
the correlation functions used to measure cosmic shear and cluster-galaxy or galaxy-galaxy lensing,
and thereby estimate how the systematic errors affect inferences about dark matter and dark energy.
The simulations can also be
used (perhaps even more importantly) for tests of mitigation schemes.

\acknowledgments

RM is supported by the Department of Energy Early Career Award Program.

\bibliographystyle{JHEP}
\bibliography{papers}

\end{document}